\documentclass[aip,apl,amsmath,amssymb,reprint]{revtex4-2}

\usepackage{graphicx}
\usepackage{dcolumn}
\usepackage{bm}
\usepackage{siunitx}
\usepackage[utf8]{inputenc}
\usepackage[T1]{fontenc}
\usepackage{mathptmx}
\usepackage{etoolbox}
\usepackage{xcolor}

\makeatletter
\def\@email#1#2{%
 \endgroup
 \patchcmd{\titleblock@produce}
  {\frontmatter@RRAPformat}
  {\frontmatter@RRAPformat{\produce@RRAP{*#1\href{mailto:#2}{#2}}}\frontmatter@RRAPformat}
  {}{}
}%
\makeatother
\begin{document}
\preprint{AIP/123-QED}

\author{Matthijs H. J. de Jong}\affiliation{Department of Precision and Microsystems Engineering, Delft University of Technology, Mekelweg 2, 2628CD Delft, The Netherlands}\affiliation{Kavli Institute of Nanoscience, Department of Quantum Nanoscience, Delft University of Technology, Lorentzweg 1, 2628CJ Delft, The Netherlands}

\author{Malte A. ten Wolde}
\affiliation{Department of Precision and Microsystems Engineering, Delft University of Technology, Mekelweg 2, 2628CD Delft, The Netherlands}

\author{Andrea Cupertino}
\affiliation{Department of Precision and Microsystems Engineering, Delft University of Technology, Mekelweg 2, 2628CD Delft, The Netherlands}

\author{Simon Gr\"oblacher}
\affiliation{Kavli Institute of Nanoscience, Department of Quantum Nanoscience, Delft University of Technology, Lorentzweg 1, 2628CJ Delft, The Netherlands}

\author{Peter G. Steeneken}
\affiliation{Department of Precision and Microsystems Engineering, Delft University of Technology, Mekelweg 2, 2628CD Delft, The Netherlands}
\affiliation{Kavli Institute of Nanoscience, Department of Quantum Nanoscience, Delft University of Technology, Lorentzweg 1, 2628CJ Delft, The Netherlands}

\author{Richard A. Norte}
\email{r.a.norte@tudelft.nl}
\affiliation{Department of Precision and Microsystems Engineering, Delft University of Technology, Mekelweg 2, 2628CD Delft, The Netherlands}
\affiliation{Kavli Institute of Nanoscience, Department of Quantum Nanoscience, Delft University of Technology, Lorentzweg 1, 2628CJ Delft, The Netherlands}

\title[Mechanical dissipation by substrate-mode coupling in SiN resonators]{Mechanical dissipation by substrate-mode coupling in SiN resonators}
\date{\today}

\begin{abstract}State-of-the-art nanomechanical resonators are heralded as a central component for next-generation clocks, filters, resonant sensors, and quantum technologies. To practically build these technologies will require monolithic integration of microchips, resonators, and readout systems. While it is widely seen that mounting microchip substrates into a system can greatly impact the performance of high-Q resonators, a systematic study has remained elusive, owing to the variety of physical processes and factors that influence the dissipation. Here, we analytically analyze a mechanism by which substrates couple to resonators manufactured on them, and experimentally demonstrate that this coupling can increase the mechanical dissipation of nanomechanical resonators when resonance frequencies of resonator and substrate coincide. More generally, we then show that a similar coupling mechanism can exist between two adjacent resonators. Since the substrate-mode coupling mechanism strongly depends on both the resonator position on the substrate and the mounting of the substrate, this work provides key design guidelines for high-precision nanomechanical technologies.
\end{abstract}
\maketitle

Optomechanics\cite{Aspelmeyer2014} represents one of the core research directions for improving the precision and accuracy of sensors, by combining the low loss of mechanical sensors\cite{Krause2012,Bagci2014,Chien2018,Gruber2019,Halg2021} with the accuracy of optical readout and control\cite{Szorkovszky2011,Purdy2013,Wilson2015}. An important figure of merit for maximizing performance is the mechanical Q-factor, which greatly reduces the effect of thermomechanical noise that limits sensors, but when considering future applications of these resonators, their footprint, fabrication complexity, and integration with other sensor components, such as the substrate, are also crucial properties. High-stress silicon nitride resonators (Si$_3$N$_4$) exhibiting state-of-the-art mechanical quality factors can be negatively impacted by interactions with their substrates. Phononic shields\cite{Yu2014,Tsaturyan2017,Ghadimi2018,Guo2019,MacCabe2020} have been used to reduce these interactions and reach exceptionally high Q-factors ($10^9$), but their size, complexity and thermal performance limits many real-world applications. It is well-known that thin and clamped-down substrates can produce significant losses in high-Q Si$_3$N$_4$ resonators\cite{Norte2016,wilson2012cavity}, but to date, little is known about their precise interaction. Several works have focused on acoustical impedance mismatching or phonon tunneling\cite{WilsonRae2008,WilsonRae2011,Rieger2014} to study and minimize dissipation channels of mechanical resonators to their environment by treating the substrate as a semi-infinite structure, and some works have studied the interaction between resonator modes and the substrate\cite{Joeckel2011,Chakram2014,Borrielli2016,Miller2021}. In this work, we build on this latter direction by linking it to the well-known effect of dissipation in resonant coupled resonators~\cite{Dolfo2018}, and show that coupling between resonator and substrate modes can negatively affect the Q-factor of trampoline resonators\cite{Norte2016} despite their difference in size. We deliberately fabricate resonators with resonance frequencies near those of a substrate mode, and show that their dissipation is increased by the coupling to this low-Q substrate mode. Furthermore, we show that the substrate can even mediate resonant coupling between two resonators separated by \SI{1.5}{\milli\meter}, which can provide an additional loss path when the density of resonators on a microchip is increased. With this study, we show the mechanism by which resonators and substrate couple and highlight the largely unexplored effect of substrate design, which can prove to be important for future optomechanical microchip designs, particularly when considering arrays of high-Q mechanical resonators\cite{Li2018,Westerveld2021}.

\begin{figure}
\includegraphics[width = 0.5\textwidth]{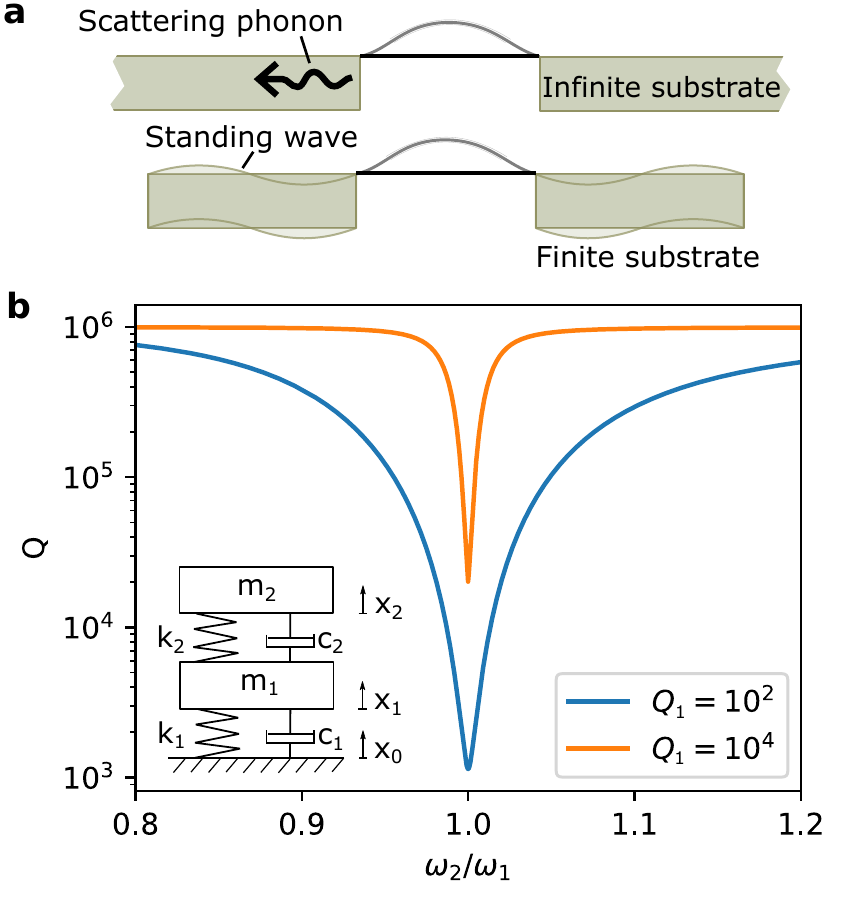}
\caption{\textbf{a}: Schematic of (incoherent) phonon scattering into an infinite substrate (top), and (coherent) phonon transfer into a discrete mode of a finite substrate that is the focus of this work (bottom). \textbf{b}: Mode coupling between a high-Q resonator ($m_2$) and a low-Q substrate mode ($m_1$, inset) reduces the effective resonator Q (y-axis) if their frequencies $\omega_1$ and $\omega_2$ are identical. The reduction of resonator Q depends on the intrinsic substrate $Q_1$ as indicated by the difference between the orange ($Q_1 = 10^4$) and blue ($Q_1 = 10^2$) curves.}
\label{FigAnalyitcalmodel}
\end{figure}

The substrate, to which high-tension Si$_3$N$_4$ membranes are anchored, is often treated as a fixed boundary (i.e.~a simple spring model)\cite{Norte2016,Gao2020,Hoj2021}. This simplification results in a negligible error when considering the mode shapes and frequencies of the resonators, since the stiffness and mass of the (typically $\sim$\SI{500}{\micro\meter}) thick substrate are much bigger than that of the thin membrane. Through the mode shape and frequencies, the fixed-boundary method correctly takes into account bending (and intrinsic) losses\cite{Schmid2011,Shin2022}, and by adding a lossy spring model, one can take into account radiative losses to traveling waves in the substrate\cite{WilsonRae2008,Tsaturyan2014,Gao2020} (phonon tunneling, cf. Fig.~\ref{FigAnalyitcalmodel}\textbf{a}, top) as well. However, this method does not treat losses due to coupling to a specific substrate \emph{resonance mode} (Fig.~\ref{FigAnalyitcalmodel}\textbf{a}, bottom), which might reduce Q when particular modes of the resonator and substrate coincide, an effect well-known from classical mechanics\cite{Dolfo2018}. 

To gain insight, we consider a simple analytical model of two stacked and coupled masses $m_1,~m_2$ with springs $k_1,~k_2$ and dampers $c_1,~c_2$ (see inset of Fig.~\ref{FigAnalyitcalmodel}\textbf{b}), representing a light resonator coupled to a heavy substrate ($m_2 \ll m_1$). Without driving, the equation of motion describing the positions of the masses $x_1,~x_2$ for this system is 
\begin{equation}
\begin{bmatrix}
(k_1 - \omega^2 m_1) + i\omega c_1 & -m_2 \omega^2 \\
-(k_2 + i \omega c_2) & (k_2 - \omega^2 m_2) + i \omega c_2
\end{bmatrix}
\begin{bmatrix}
x_1 \\
x_2
\end{bmatrix}
= 0
\label{EOM}
\end{equation}
which we can straightforwardly solve for complex eigenfrequencies $\omega_i$ ($i= 1,2$) from which we can extract the Q-factor via 
\begin{equation}
Q_i = \frac{\mathrm{Re}(\omega_i)}{2 \mathrm{Im}(\omega_i)}.
\end{equation}
We use realistic parameters $m_1=$~\SI{1.47}{\milli\gram} and $m_2=$~\SI{11.8}{\nano\gram} for the effective masses\cite{Hauer2013} of substrate and resonator mode, choose $\omega_1 = \sqrt{k_1/m_1} =2\pi\times$\SI{100}{\kilo\hertz}, and choose $c_1,~c_2$ such that our resonator is intrinsically limited to $Q_2=10^6$ but our substrate $Q_1$ is substantially lower. Then we vary $\omega_2$ by adjusting $k_2$. When the (real part of the) eigenfrequencies of the two modes is very different ($\omega_2 \neq \omega_1$), the two resonances are essentially independent; thus, there is little energy transfer between the modes. However, when their eigenfrequencies are closer together ($\omega_2 \approx \omega_1$), the modes hybridize and energy transfer from one mode to the other can occur\cite{Dolfo2018,Zanette2018}. If the damping of the substrate mode is higher than that of the resonator mode, the substrate mode essentially functions as an additional loss mechanism for the resonator mode as shown in Fig.~\ref{FigAnalyitcalmodel}\textbf{b}. The frequency range over which the energy transfer is significant is determined both by the Q-factor of the low-Q mode and by the difference in mass/stiffness of the two resonators. From this basic model, it is expected that low-Q substrate modes might have significant impact on the Q-factor of high-Q resonators under certain conditions. We will numerically and experimentally explore this loss mechanism in more detail for Si$_3$N$_4$ trampoline resonators. 

We use a finite-element model of our resonator and substrate to numerically analyze the loss mechanism by substrate-resonator mode-coupling (see Supplementary Information (SI) Sec.~S1 for details). We take a viscoelastic material loss model for both the substrate\cite{Schmid2011,Joeckel2011} (loss factor $\eta_\mathrm{Si} = 10^{-4}$) and membrane\cite{Fedorov2018} ($\eta_\mathrm{SiN}= 10^{-7}$), where we choose the values such that $Q = \eta^{-1}$ matches with experimental observations of the substrate modes (SI Sec.~S2) and resonator modes respectively. To distinguish these Q-factors, we will refer to the viscoelastically-limited (intrinsic) Q's as $Q^i$, and the hybridized Q's with $Q^h$. 

\begin{figure*}
\includegraphics[width = \textwidth]{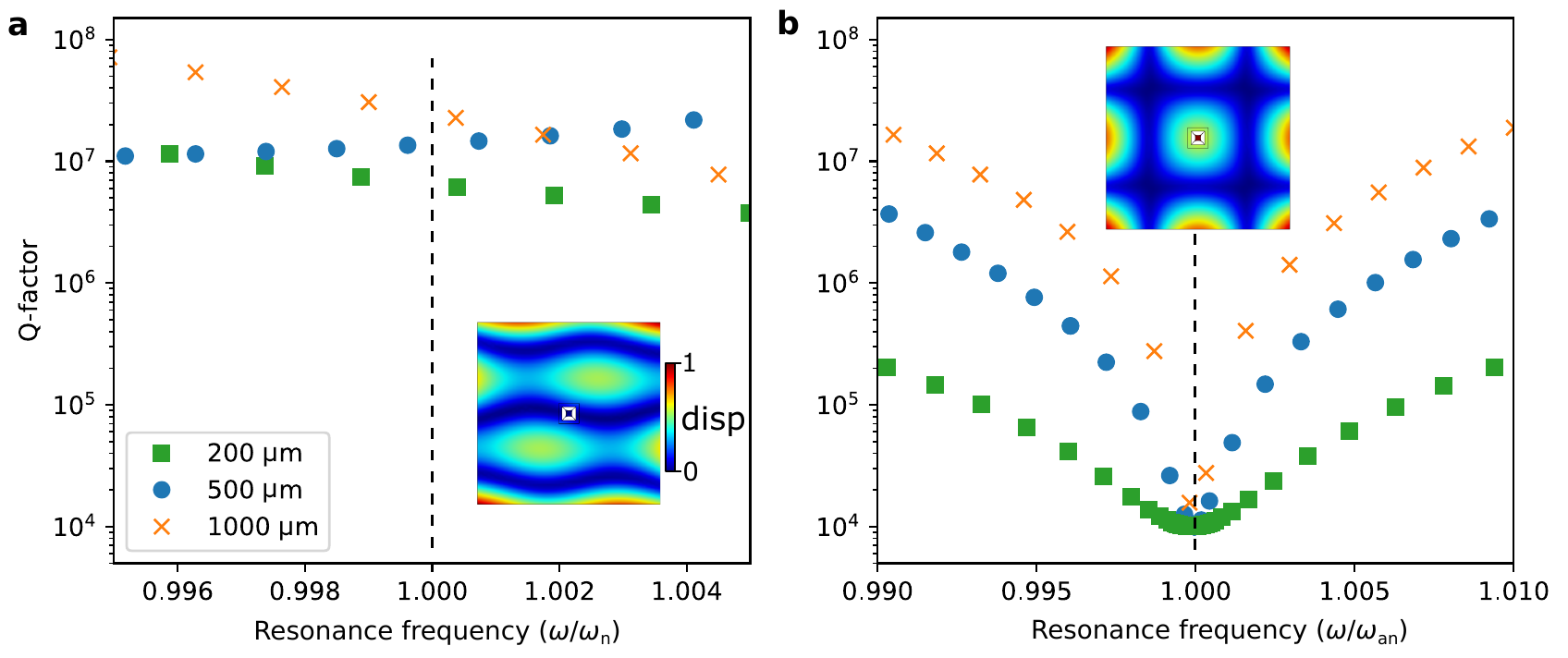}
\caption{Simulated Q-factor $Q^h_2$ of resonator mode at $\omega_2$, at different values of $\omega_2 \approx \omega_1$ coupling to two different substrate modes, with \textbf{a} a node at the resonator position ($\omega_1 = \omega_n$), and \textbf{b} an anti-node at the resonator position ($\omega_1 = \omega_{an}$). Insets show the normalized out-of-plane displacement of the substrate mode, with the resonator located in the center. The lossy substrate mode significantly reduces the Q-factor of the resonator mode over a large frequency range when located at an anti-node (\textbf{b}), but has little effect when located at a node (\textbf{a}). Different curves show the effect for different substrate thicknesses.
}
\label{figure1}
\end{figure*}

In Fig.~\ref{figure1}, we plot the Q-factor of the simulated membrane mode as a function of resonator mass, such that its resonance frequency crosses two substrate modes. When the resonator frequencies are very different, the resonator's Q is limited by the Si$_3$N$_4$ material loss, $1/\eta_\mathrm{SiN} \simeq 10^7$ so $Q^h_2 = Q^i_2$, as expected from uncoupled modes. Close to a substrate mode (dashed line), the Q-factors of the modes hybridize similarly to the analytical model; $Q^h_2$ decreases to $Q^h_1 \simeq 1/\eta_\mathrm{Si} \simeq 10^4$ limited by the substrate material loss. Here, energy-loss via coupling to the lossy substrate mode is the dominant loss mechanism. Not all substrate modes decrease the resonator $Q^h_2$ equally, e.g.~the mode of Fig.~\ref{figure1}\textbf{a} with frequency $\omega_1 = \omega_\mathrm{n}$ shows no decrease, while the mode of Fig.~\ref{figure1}\textbf{b} with frequency $\omega_1 = \omega_\mathrm{an}$ shows a pronounced decrease. If the resonator is located at a node of the substrate mode (mode shape shown in Fig.~\ref{figure1} insets), there is no motion to couple to, so there is almost no energy transfer between the modes. Trends visible in the resonator Q-factor in Fig.~\ref{figure1}\textbf{a} are attributed to nearby substrate modes (not shown) that do couple to the resonator mode.

Aside from the mode shape, the substrate thickness also affects the mode coupling, as can be seen from the different colored curves of Fig.~\ref{figure1}. While $Q^h_2$ goes to the same level when the resonance frequencies are equal (if $\omega_2 = \omega_1$, $Q^h_2 \rightarrow Q^h_1 \simeq Q^i_1 \simeq 1/\eta_\mathrm{Si} \simeq 1\times 10^4$), the frequency range over which this happens is much more narrow for a thick substrate. The reason is that the mass difference between resonator and substrate is bigger for a thicker substrate, which reduces the effective coupling between the masses (top-right term in Eq.~\eqref{EOM} after normalization). While the shape and size of a substrate are important parameters for the frequency distribution of substrate modes, Eq.~\eqref{EOM} and the results of Fig.~\ref{figure1} suggest that the substrate mass governs the coupling strength between resonator and substrate modes at resonance. This points to thicker substrates being better (less coupled) in general, but for a given substrate thickness and resonator frequency, the optimal substrate shape and size must be carefully designed.

\begin{figure*}
\includegraphics[width = \textwidth]{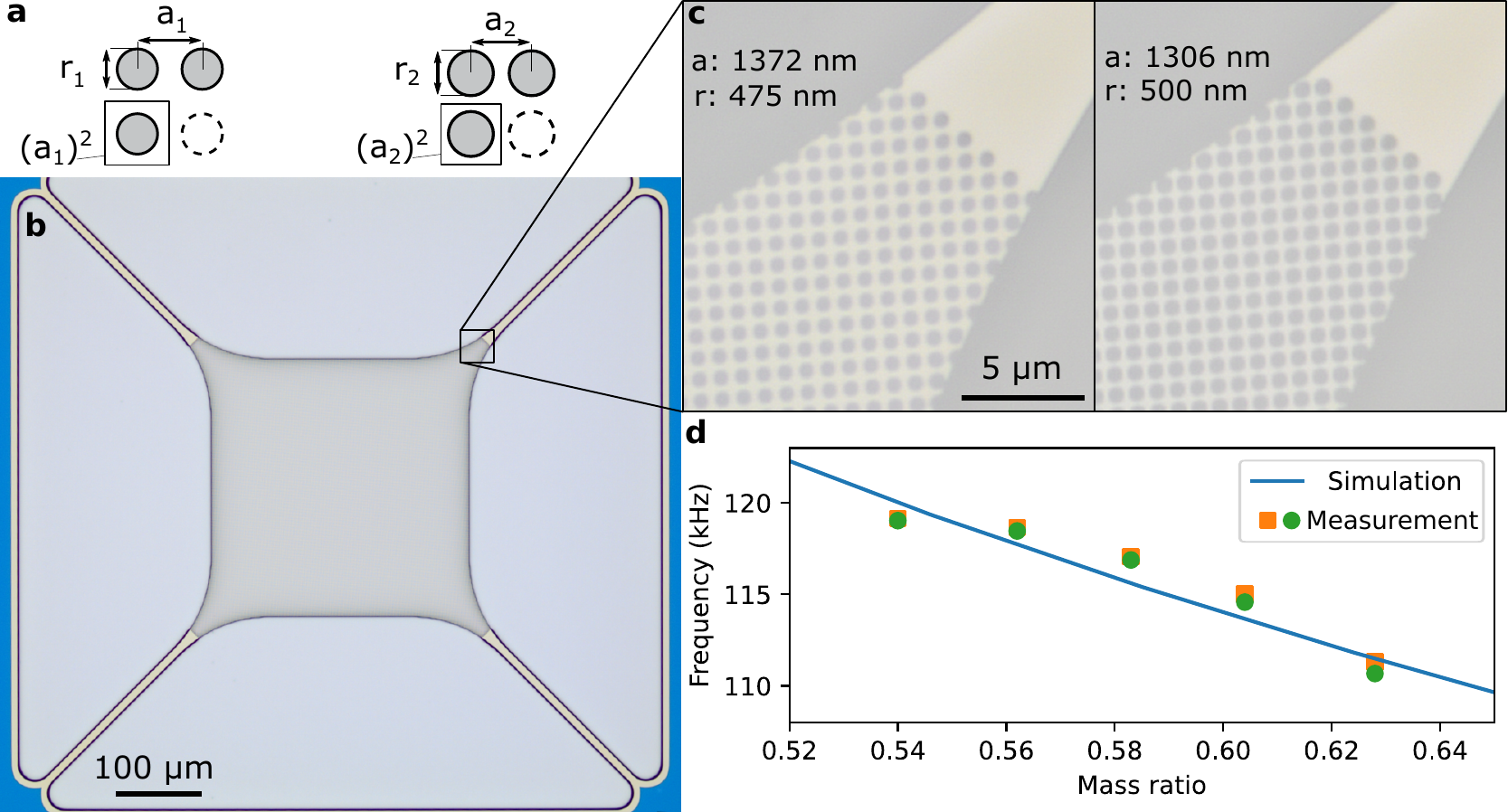}
\caption{\textbf{a}: Schematic of variation of photonic crystal parameters used to change the mass ratio. \textbf{b}: Optical microscope image of suspended membrane, the blue color is from thin-film interference effects of the SiN. \textbf{c}: Zoom-in of photonic crystal edge to show change in hole size and spacing. \textbf{d}: Simulated and measured resonator frequencies as function of designed $r_\mathrm{m}$ for two nominally identical chips. Standard deviation of frequencies is approximately equal to the size of the data points.
}
\label{figure_PhC_sweep}
\end{figure*}

To investigate the effect of coupling to the substrate mode on the resonator's Q-factor $Q^h_2$, we fabricate (see SI Sec.~S3) resonators with slightly different resonance frequencies, by varying the membrane's mass-per-area by perforating it using small holes of controlled radius. This square lattice of holes also functions as a photonic crystal to increase the membrane's reflectivity\cite{Gaertner2018} and causes the membrane to release evenly during the fabrication process. Since this method ensures that the geometry of the resonator is almost constant, this allows varying the resonance frequency with minimal effect on the Q-factor\cite{Norte2016,Bereyhi2019,Sadeghi2019}. We change lattice constant $a$ and hole radius $r$ (Fig.~\ref{figure_PhC_sweep}\textbf{a-c}). The mass ratio $r_\mathrm{m} = 1 - \pi r^2/a^2$ relates the mass of the patterned photonic crystal to the mass of unpatterned Si$_3$N$_4$. The range over which $r_\mathrm{m}$ can be varied is limited, due to stress focusing (see SI Sec. S4 for details) and fabrication constraints. The effect of $r_\mathrm{m}$ on $Q^i_2$ is negligible; simulations predict at most 20\% change over the parameter range (SI Sec. S4), confirmed by the absence of a dependence of the measured values of $Q^h_2$ on the photonic crystal parameters. We use $10\times 10 \times 1$~mm$^3$ chips with 25 membranes fabricated with five different $r_\mathrm{m}$. The measured resonance frequencies of their fundamental modes agree well with simulations (Fig.~\ref{figure_PhC_sweep}\textbf{d}). 

\begin{figure*}
\includegraphics[width = \textwidth]{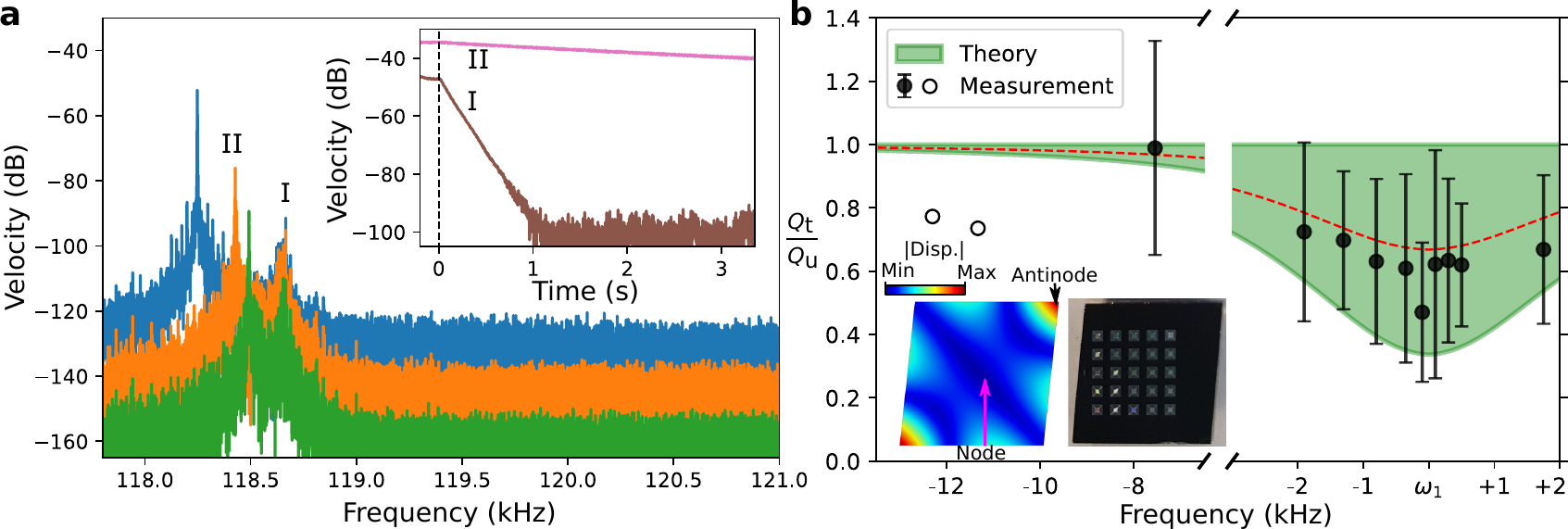}
\caption{\textbf{a}: Mechanical velocity spectrum ($20\log_{10}(v/v_\mathrm{ref})$ with $v_\mathrm{ref} = 1$~\si{\meter\per\second}) of three different devices, showing fundamental mode (II) close to the substrate mode (I) by driving with white noise. Inset: Ringdowns of the untaped device fundamental mode (II) and substrate mode (I), showing the difference in their Q's. \textbf{b}: Ratio of $Q^h_2$ measured on a taped ($Q_\mathrm{t}$) versus untaped ($Q_\mathrm{u}$) substrate: the increase in substrate losses causes a decrease in $Q^h_2$ when the modes are close in frequency. Theory curve shows expected reduction in $Q^h_2$ around $\omega_1$ for $Q^i_1 = 1.2\times 10^4 \rightarrow 3\times 10^3$ when applying tape. Insets shows the simulated mode shape of the substrate mode and a photo of the fabricated chip with 25 devices.}
\label{FigModecoupling}
\end{figure*}

We utilize a Polytec MSA400 laser Doppler vibrometer to spatially resolve mode shapes, obtain resonance spectra, and to acquire time traces from which we extract $Q^h_2$ via ringdown measurements (see SI Sec.~S3). In Fig.~\ref{FigModecoupling}\textbf{a}, we show the mechanical spectrum for three trampoline resonators with nominally the same $r_\mathrm{m} = 0.54$, where the fundamental mode (II) is close to a substrate mode (I). The spread in frequency due to fabrication imperfections is $<300$~\si{\hertz} on $115$~\si{\kilo\hertz}, which highlights our control over the mechanical frequencies. The inset shows for a particular device the ringdowns of the membrane ($Q^h_2 = 1.2\times 10^6$) and substrate modes.

There is a spread in resonator $Q^i_2$ (see SI Sec.~S4 and Fig.~S5) that could obscure an absolute reduction of resonator $Q^i_2$ due to coupling to the substrate mode. We can isolate the effect of the substrate coupling by controlling the substrate Q-factor $Q^i_1$. By adding carbon tape between substrate and stainless steel sample holder (see SI Sec.~S3 for the measurement protocol), we reduce\cite{Schmid2011,Tsaturyan2014} $Q^i_1$ from $1.2 \times 10^4$ (resting without tape) to $\sim 3\times 10^3$ (with tape, see SI Sec.~S2). By comparing the resonator's hybridized Q-factor $Q^h_2$ for an untaped chip ($Q_\mathrm{u}$) to the resonator's hybridized Q-factor for a taped chip ($Q_\mathrm{t}$), we isolate the effect of the substrate-mode coupling. That is, the ratio $Q_\mathrm{t}/Q_\mathrm{u}$ should be smaller than one only due to the enhanced dissipation by mode coupling.

We plot the ratio $Q_\mathrm{t}/Q_\mathrm{u}$ for 152 measurements from the resonators spread over four chips in Fig.~\ref{FigModecoupling}\textbf{b}. $Q_\mathrm{t}$ and $Q_\mathrm{u}$ are each determined by the average of three ringdowns on the same device. The data are then binned by frequency with respect to the substrate mode, for each bin we determine the mean and standard deviation to obtain the errorbars. Circles indicate single devices. Fig.~\ref{FigModecoupling}\textbf{b} also shows the theoretical analytical model introduced by Eq.~\ref{EOM}, the upper bound corresponds a membrane located at a node and thus not coupled, while the lower bound corresponds to a membrane located at an antinode, maximally coupled (simulated mode shape inset in Fig.~\ref{FigModecoupling}\textbf{b}). The red dotted line indicates the expected mean reduction in Q-factor.

Close to the substrate mode at $\omega_1$, the average $Q_\mathrm{t}/Q_\mathrm{u}$ is reduced, and closely matches the theoretical mean, while far away from $\omega_1$ it is close to 1. To gauge the statistical significance of the reduction of $Q_\mathrm{t}/Q_\mathrm{u}$, we perform Welch's t-test on the mean Q-factor ratios close to the substrate mode ($\omega_1 - 2 \text{kHz}<\omega_2<\omega_1 + 2 \text{kHz}$) and far away from the substrate mode ($\omega_2 < \omega_1 - 7 \text{kHz}$). This tests our hypothesis (Q-factor reduced close to $\omega_2$) against the null hypothesis (Q-factor not affected by $\omega_2$). We obtain a probability $p = 0.00072$, so we can reject the null hypothesis. This means the reduction in Q-factor close to the substrate mode is statistically significant. Additionally, the spread in the measured ratio of $Q_\mathrm{t}/Q_\mathrm{u}$ can be attributed predominantly to the positioning of the resonators on the chip with respect to the nodes or antinodes of the substrate mode (inset of Fig.~\ref{FigModecoupling}\textbf{b}). This effect is illustrated by the green shaded area bounded by theory. In some resonators, there is heating and optothermal driving from the laser (\SI{1}{\milli\watt} continuous-wave power) which affects the ringdown measurement, and we have excluded these devices (see SI Sec.~S5, and Fig.~S5). Summarizing, we find a significant reduction of the average $Q_\mathrm{t}/Q_\mathrm{u}$ close to the substrate mode $\omega_1$, which quantitatively agrees with the theoretical model of substrate-mode coupling, thus supporting the hypothesis that coupling to the substrate increases dissipation of the membrane mode. 

\begin{figure*}
\includegraphics[width = \textwidth]{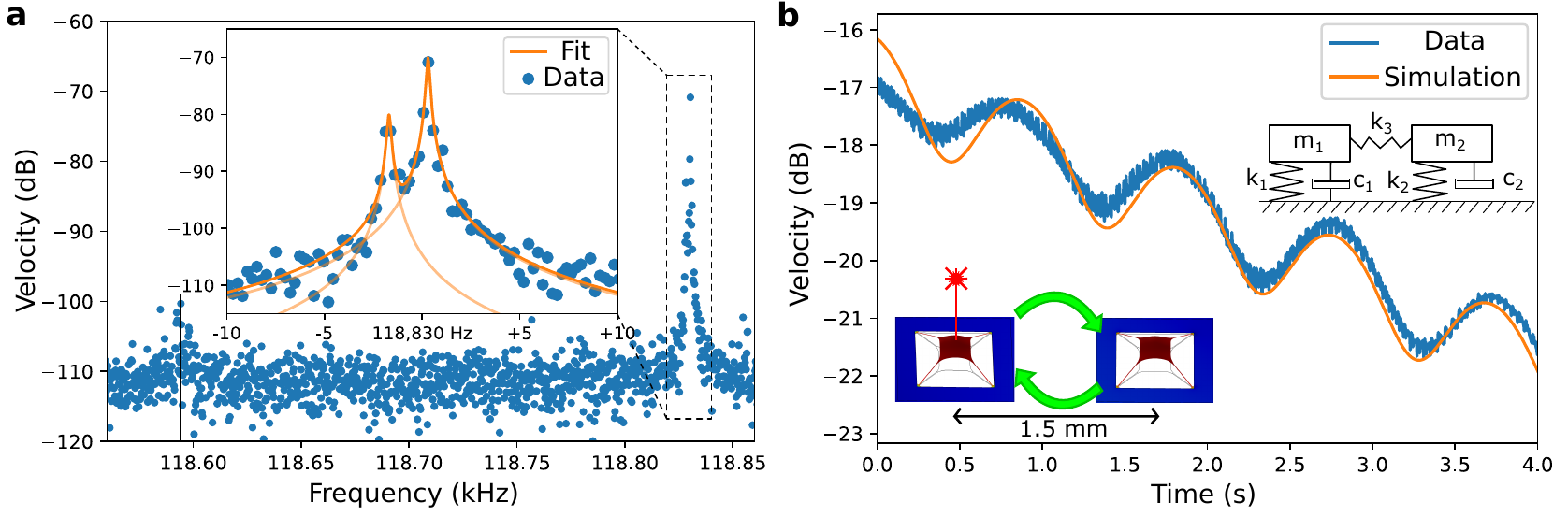}
\caption{\textbf{a}: Spectrum measured on one membrane containing both a signature of the substrate mode (black bar) and of a second membrane extremely close in frequency. Inset shows membrane peaks fitted with two Lorentzians (orange, with semi-transparent the separate Lorentzians). \textbf{b}: Ringdown (blue) by driving at the resonance of one membrane and recording the time-trace of that same membrane. Oscillations are due to coherent coupling between the two resonators spaced by \SI{1.5}{mm} (bottom inset). Simulated ringdown (orange) of two coupled resonators (top inset) using fit parameters obtained from \textbf{a}.}
\label{FigResonatorcoupling}
\end{figure*}

After having investigated the importance of resonator-substrate coupling, we now address the possibility of two resonators on the same chip affecting each other. Such couplings can be relevant in resonator arrays, and have been found in lower-Q devices\cite{Luo2018,Zhang2020,Siskins2021}. By measuring their resonance frequencies, we identify two membranes spaced \SI{1.5}{\milli\meter} apart (see SI Sec. S6 for details) with resonance frequencies identical to within \SI{2}{\hertz} ($\omega_1/2\pi=$~\SI{118.828}{\kilo\hertz} and $\omega_2/2\pi=$~\SI{118.830}{\kilo\hertz}), much closer together than either of them are to the substrate mode, Fig.~\ref{FigResonatorcoupling}\textbf{a}. From Lorentzian fits to the spectrum (orange curves), we extract their Q-factors, $Q_1 \simeq 0.6\cdot 10^6$ and $Q_2 \simeq 0.8\cdot 10^6$.

By driving at the resonance of one membrane and recording the ringdown, we see oscillatory behavior which we model by two discrete coupled resonators, Fig.~\ref{FigResonatorcoupling}\textbf{b} and SI Sec. S6 for details. The equation of motion for the resonator positions is 
\begin{equation}
\begin{bmatrix}
(k_1 - \omega^2 m_1) + i\omega c_1 & J^2 \\
J^2 & (k_2 - \omega^2 m_2) + i\omega c_2 
\end{bmatrix}
\begin{bmatrix}
x_1 \\
x_2 
\end{bmatrix}
= 0,
\label{Coupledmodel}
\end{equation}
where the coupling between the resonators via the substrate is modeled by the parameter $J$. The indices 1,2 now both refer to the two membranes, and $J^2 = \frac{k_3^3}{m_1 m_2}$ is the coupling rate between them (Fig.~\ref{FigResonatorcoupling}\textbf{b}, inset). By integrating the equations of motion, Eq.~\ref{Coupledmodel}, and plotting the resulting velocity of one of the resonators, we can nearly exactly reproduce the oscillating ringdowns we observe after having adjusted the initial position to get a good fit. The oscillations in the ringdown can be attributed to energy exchange between the spatially separated resonators through the substrate. Based on the periodicity, we extract a coupling rate $J/2\pi \simeq 138$~\si{\hertz}. When a linear fit is made though the middle of the oscillations, we obtain a Q-factor of $Q_\mathrm{tot} = 0.83\cdot 10^6$, corresponding reasonably well to the Q-factors from the Lorentzian fits. This measurement demonstrates that the on-chip coupling between Si$_3$N$_4$ membranes on the same substrate can present an important coupling channel. Furthermore, the oscillating behavior implies coherence in the energy exchange, which is of interest for information processing\cite{Faust2013,Luo2018,Zhang2020,Siskins2021} in particular if a control mechanism to adjust the coupling can be devised. The fact that we see such energy exchange in a passive system suggests it should be taken into account when designing sensors based on resonator arrays\cite{Li2018,Westerveld2021}.

In conclusion, we demonstrate analytically, numerically, and experimentally a mechanism behind the coupling between high-Q resonators and substrate modes, which can reduce the Q-factor of the resonators when their frequencies match. Using a laser Doppler vibrometer to identify resonator and substrate modes, we are able to explain the physics behind this interaction. Interestingly, this interaction is not only limited to resonator and substrate but also exists between spatially separated high-Q resonators under the same frequency-matching condition. These behaviors in a fully passive system show the importance of considering resonator-substrate interactions in future designs of arrays of high-Q mechanical resonators for sensing, actuation, filtering and timing applications. In particular, thin and clamped-down substrates may have a dense spectrum of low-Q modes and suffer from resonator-substrate interaction as a result. To avoid these interactions, our numerical results point toward thick substrates for their increase in mass and stiffness\cite{Norte2016}, and laterally small chips for a sparser spectrum of substrate modes. We further confirm the result that avoiding tape to mount chips to a sample holder is best to retain high resonator Q-factors. Neither of these effects had been systematically explored before due to the stringent requirements on resonator frequency precision. This work, thus, highlights the substrate and mounting as important parameters to incorporate in future design methodologies.

\section*{Author Contributions}
M.W.~and P.S.~performed the analytical calculations, M.J.~and M.W.~performed the finite-element modeling. M.J.~and M.W.~conducted the experiment, A.C.~developed the fabrication process and fabricated the devices. S.G.~is co-supervisor to M.J. M.J.~analyzed the experimental data. R.N.~designed the research and supervised the project. All authors contributed to the writing of the paper.

\section*{Supplementary Materials}
See supplementary material for details regarding sample fabrication, measurement method and simulation, as well as additional measurements and simulations supporting the main text.

\section*{Acknowledgements}
The authors want to thank Dongil Shin for his help in simulations and Lauren Vierhoven for initial experimental help. We also acknowledge Kavli Nanolab Delft for fabrication assistance. The research leading to these results has received funding from the European Union’s Horizon 2020 research and innovation programme under Grant Agreement Nos. 785219 and 881603 Graphene Flagship. This work has received funding from the EMPIR programme co-financed by the Participating States and from the European Union’s Horizon 2020 research and innovation programme (No. 17FUN05 PhotoQuant). 

\section*{Data availability}
The data that support the findings of this study are openly available in  the 4TU Research Data Repository at \href{https://doi.org/10.4121/19209333.v2}{https://doi.org/10.4121/19209333.v2}


\setcounter{figure}{0}
\renewcommand{\thefigure}{S\arabic{figure}}
\setcounter{equation}{0}
\renewcommand{\theequation}{S\arabic{equation}}

\clearpage
\section{Membrane-on-substrate simulations}
\begin{figure}
\includegraphics[width = 0.5\textwidth]{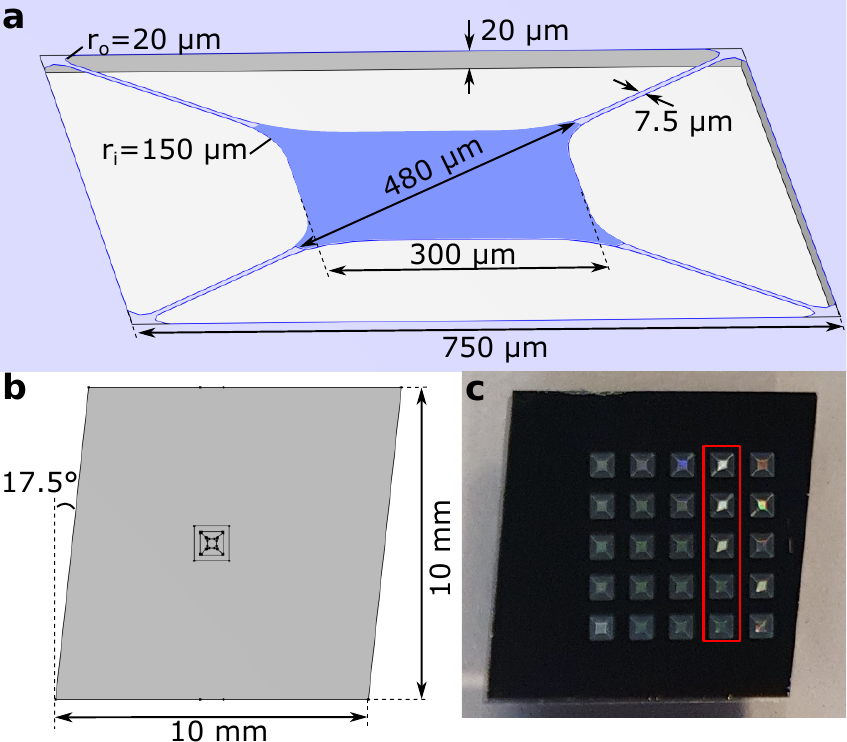}
\caption{\textbf{a:} Simulation setup of Si$_3$N$_4$ membrane as a shell (light purple/blue) on top of a Si solid (grey), with nominal membrane design parameters. \textbf{b:} Full simulation domain, matching closely in size to the actual chips. \textbf{c:} Image of chip containing 25 membranes with 5 different designs, one for each column (red).}
\label{figure_simulation_setup}
\end{figure}

We use a COMSOL\textsuperscript{\small{\textregistered}} model to simulate mode frequencies and shapes, and obtain estimates of the Q-factor of the different resonances. The model consists of a 2D shell (Si$_3$N$_4$) and a 3D solid (Si) with a solid-shell connection representing the chemical bond between the two, as in Fig.~\ref{figure_simulation_setup}. Around the patterned membrane, there is a \SI{20}{\micro\meter} cutout in the Si such that the Si$_3$N$_4$ is suspended. Upon release, the \SI{1}{\giga\pascal} pre-stress in the Si$_3$N$_4$ redistributes (described in the next section), to take this into account we first perform a stationary step before calculating the eigenmodes of the system. This also accounts for the membrane geometry changing due to the stress redistribution, though this effect is minimal. 

The membrane geometry, Fig.~\ref{figure_simulation_setup}\textbf{a}, was designed in a different work~\cite{Norte2016}, though slight modifications were made to account for a new etch process. The diameter of the photonic crystal (\SI{480}{\micro\meter}) extends significantly beyond the width of the membrane pad (\SI{300}{\micro\meter}). The photonic crystal holes also function as etch release holes, which is crucial to avoid the membrane collapsing. As the tether width increases towards the pad (start at \SI{7.5}{\micro\meter}, with fillet radii of \SI{150}{\micro\meter}), this area is the last to be released. To increase the yield of fabrication, the photonic crystal was extended to cover this area and provide a more equal release of the membrane.

When calculating the eigenmodes of this model, we obtain the free-free modes and discard the rigid-body modes. In the experiment, the chip is placed on a stainless steel sample plate, which constrains the chip motion. Taking this interface properly into account is rather involved, so we neglect this effect. Based on the good agreement in both mode frequency and mode shape between the simulations and measurements, this is a valid simplification.

For most of this work, we follow the convention of using square ($10\times 10$~\si{\milli\meter\squared}) chips, which would result in symmetric substrate mode shapes. For thick (\SI{1}{\milli\meter}) chips in particular, the targeted mode of interest has a large area of low mode amplitude in a ring around the center of the substrate (cf. Fig. 2 of the main text), which would preclude the majority of membranes from interacting with this mode. To avoid this, we diced these chips at a slight angle, Fig.~\ref{figure_simulation_setup}\textbf{b} and \textbf{c}, such that there was a substrate mode of the right frequency with a reasonably flat mode profile across the chip. 

In the simulations supporting Fig. 2 of the main text, we sweep the resonator frequency by changing the mass of the resonator. The central pad of the trampoline membrane (blue in Fig.~\ref{figure_simulation_setup}) is assigned a different virtual material than the rest of the Si$_3$N$_4$ surface (light purple in Fig.~\ref{figure_simulation_setup}); we modify the material density $\rho$ to change the mass, but keep the rest of the material parameters the same. By choosing the material density correctly, we can sweep the resonator mode across any substrate mode of choice. For the different $10\times10$~\si{\milli\meter\squared} square chips of \SI{200}{\micro\meter}, \SI{500}{\micro\meter} and \SI{1}{\milli\meter} thickness, we choose modes with the same out-of-plane mode shape, which are at different frequencies for each of the chip thicknesses. This allows for direct comparison of the frequency range over which the resonator Q-factor is reduced due to coupling to the substrate mode.

\begin{figure*}
\includegraphics[width = \textwidth]{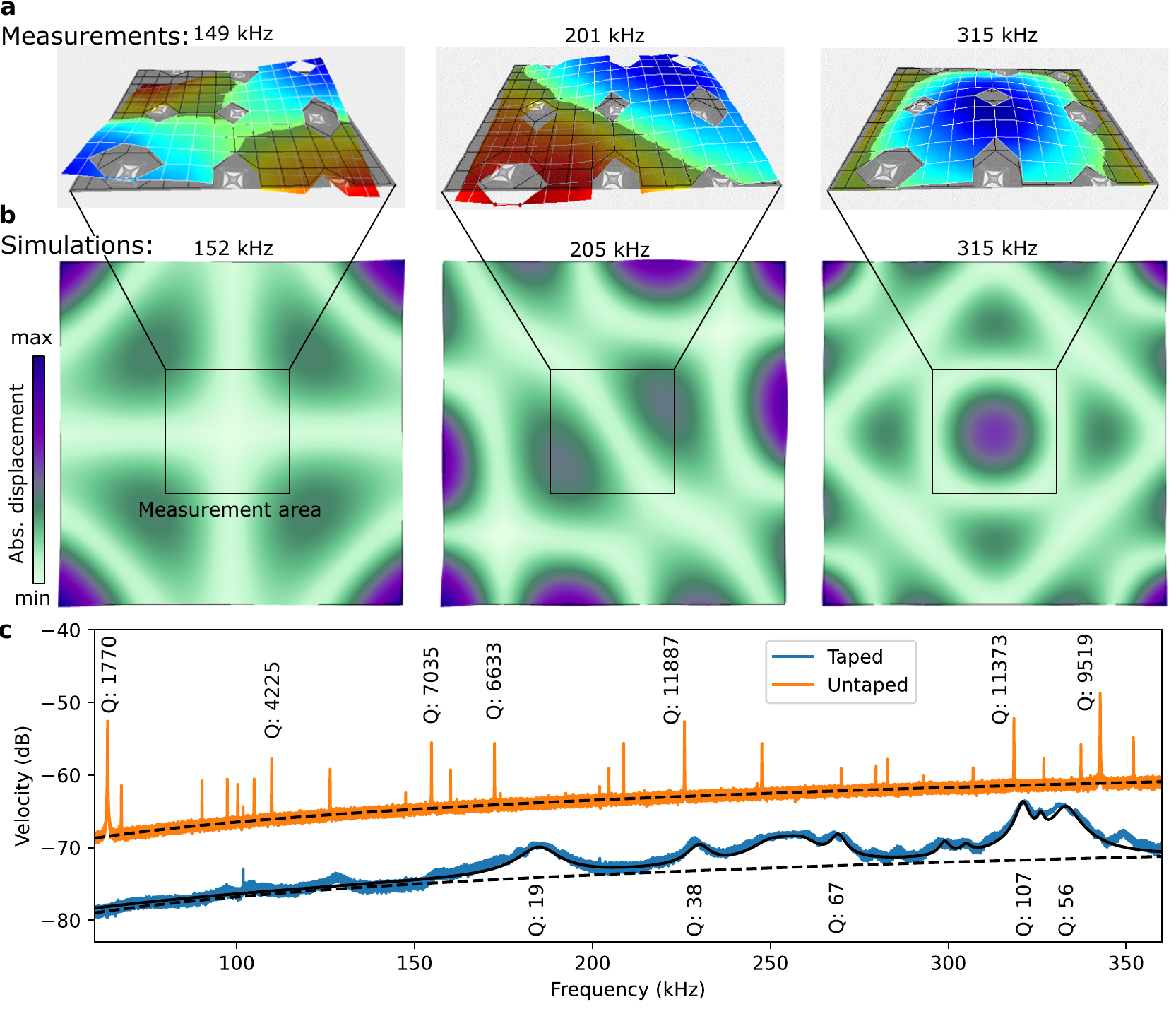}
\caption{\textbf{a:} Measured substrate modes on~\SI{200}{\micro\meter} substrates, which are easier to visualize than the substrate modes of the~\SI{1}{\milli\meter} thick substrates used in the main text. \textbf{b:} Simulated substrate modes for the~\SI{200}{\micro\meter} substrates, showing good agreement in both mode shape and frequency with the measured results. \textbf{c:} Spectrum of the measured substrate modes with and without tape, and Q-factors of prominent modes. Dashed lines indicate the detector noise floor, solid black line is Lorentzian fit (only shown for taped case). The curves are vertically offset for clarity.}
\label{figure_substrate_mode_shape}
\end{figure*}

\section{Substrate modes}
We verify the mode shapes we simulate with spatially-resolved mode measurements using the Polytec MSA400 laser Doppler vibrometer. For a nearly-perfectly square $10\times 10$~\si{\milli\meter\squared} Si chip of \SI{200}{\micro\meter} thickness, we show three modes (149, 201 and \SI{315}{\kilo\hertz}) in Fig.~\ref{figure_substrate_mode_shape}\textbf{a}. Simulations of this chip design show excellent agreement in both mode shape and frequency with the measured results, Fig.~\ref{figure_substrate_mode_shape}\textbf{b}. Due to the difference in thickness, these chips have different substrate modes that are considerably easier to measure and visualize than the \SI{1}{\milli\meter} chips used for the experiments in the main text.

To motivate the loss factors used for the modeling in the main text, we fit a Lorentzian to the mechanical modes visible in the substrate spectrum, shown in Fig.~\ref{figure_substrate_mode_shape}\textbf{c}. When the substrate is not taped to the sample holder, the modes have a Q-factor on the order of $10^4$, but that is decreased to $10^2$ when we add carbon tape. In the latter case, the modes are sufficiently broad that they overlap so we show the sum of the different Lorentzians (solid black line) on top of the detector noise floor (dashed black line) in Fig.~\ref{figure_substrate_mode_shape}\textbf{c}. For some of the more prominent modes, we have denoted the Q-factor in the figure. Note that the substrate mode Q-factors reported here are lower than the ones reported in the main text, which is due to the thickness of the chip used (\SI{200}{\micro\meter} in Fig.~\ref{figure_substrate_mode_shape}\textbf{c} versus \SI{1}{\milli\meter} in the main text).
\\

\begin{figure}
\includegraphics[width = 0.5\textwidth]{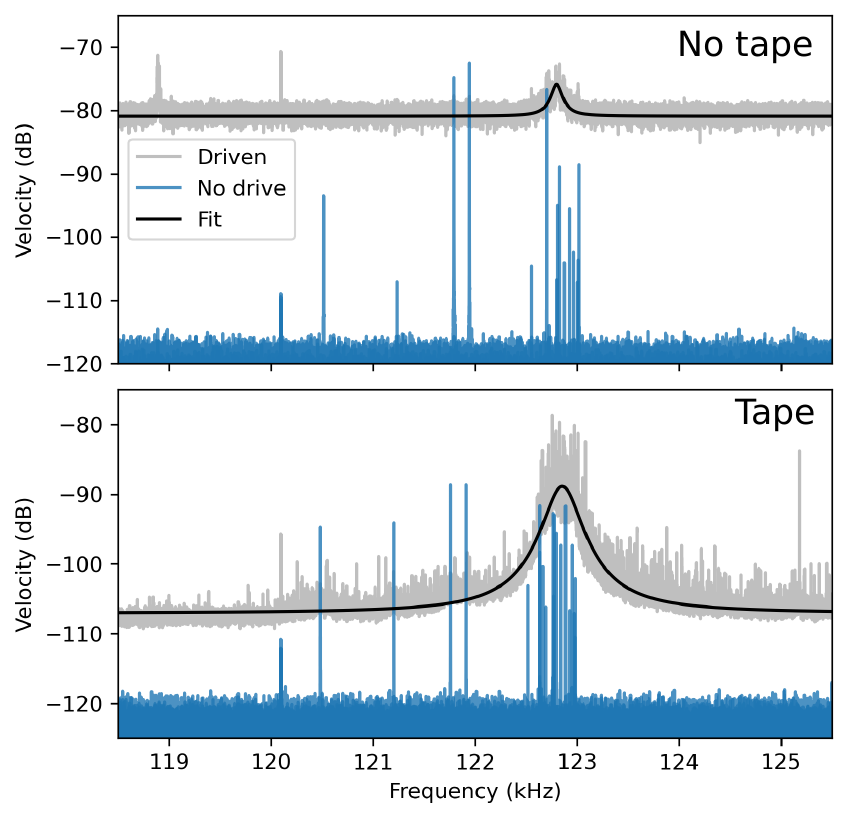}
\caption{Comparison of substrate modes without (top) and with (bottom) tape. Grey curves show the average of white-noise-driven measurements of all devices on a particular chip (offset vertically for clarity). Blue curves show all the individual measurements without drive (stacked), representing all membrane resonances. In black, Lorentzian fit to the substrate mode.}
\label{figure_substrate_mode_comparison}
\end{figure}

For a quantitative match between the analytical theory and the obseved reduction in membrane Q-factor, we must obtain the the substrate Q-factor with and without tape. However, for thicker chips the amplitude of this substrate is too small to reliably measure at any single membrane position. It can be amplified by strong driving with white noise, but obtaining a fit is made difficult by the membrane mode. However, by averaging measurements from all membranes, we can isolate the mode they have in common which should be the substrate mode. We do so in Fig.~\ref{figure_substrate_mode_comparison}, where we plot the spectra of the averaged driven measurement (grey), and the collective of measurements without driving (blue). The two are offset vertically for clarity. We obtain a Lorentzian fit at \SI{122.85}{\kilo\hertz}, with linewidths corresponding to $Q = 1.2\times 10^4$ (no tape) and $Q = 4 \times 10^3$ (tape), which are the values used in the main text.

It is worth noting that these fits come with some uncertainty, as we cannot exclude that there is some remaining signal from any of the membrane modes. These fits also represent only a single chip, though the frequencies and Q-factors of the substrate modes of the other chips used in this work are similar. Fig.~4 of the main paper combines the results of these three chips, and the confidence interval of that theory fit encompasses the spread in substrate Q-factors of the three chips.

\section{Measurement method and fabrication}
The spectra and ringdowns shown in this work were obtained using a Polytec MSA400 laser Doppler vibrometer. It uses the Doppler frequency shift of light reflected from a device under test due to the out-of-plane motion to quantify the velocity. The device mechanical spectrum can be obtained by Fourier-transforming the time signal obtained from the vibrometer, and using the scanning stage, mode shapes can be imaged for identification of the modes. To perform a ringdown measurement, we applied a (typically) \SI{10}{\milli\volt} peak-peak sine wave using a Rigol D1032Z signal generator such that we resonantly drive the motion of a particular mode of the device under test. Then we record the time trace after stopping the driving, to observe the energy decay of the driven mode. We perform a short-time Fourier transform on successive parts of the time-signal using scipy’s stft function to obtain the temporal behaviour of the mechanical spectrum. By making a line-cut along a particular frequency, we can extract peak amplitude as a function of time. In logarithmic scale, we expect a linear curve where the slope $b$ corresponds to the Q-factor via $Q = \frac{2f}{\log{10}b/10}$. This way, we extract the Q-factor. 

We performed the experiments by placing a chip containing $25$ devices (membrane) on the sample holder inside the vacuum chamber (i.e. without tape). We operate at a pressure $<1\times 10^{-5}$~\si{\milli\bar}. For every device, we record the spectrum with and without white noise driving, and determine the frequency of the fundamental mode. Then, we perform three sequential ringdown measurements at that frequency. After every device is measured, we vent the chamber and remove the chip. We then apply a piece of carbon tape to the sample holder, taking care to use a similar-size piece every time. We place the chip such that the carbon tape is in the center of the chip. To create the most repeatable connection between chip, tape and sample holder we gently press down on the outside of the chip with tweezers, to ensure good contact. We then pump down and repeat all the measurements (spectrum with and without white-noise driving, and ringdowns). We perform the data analysis and fit of the Q-factor afterwards.

The trampoline resonators were fabricated on \SI{100}{\nano\meter} thick stoichiometric silicon nitride (Si$_3$N$_4$) deposited by low-pressure chemical vapour deposition onto a \SI{1}{\milli\meter} thick silicon substrate. The pattern was first written on a positive tone resist by electron beam lithography and, after the development, transferred on the Si$_3$N$_4$ layer using ICP etching. The resist was then removed using dimethylformamide followed by two cleaning steps with piranha solution and diluted hydrofluoric acid. Finally, the trampoline resonators were released using an isotropic ICP etch with SF$_6$ at $-120$~\si{\celsius} for $30$ seconds thus completing the fabrication process.

\section{Stress redistribution due to photonic crystal}
\begin{figure}
\includegraphics[width = 0.5\textwidth]{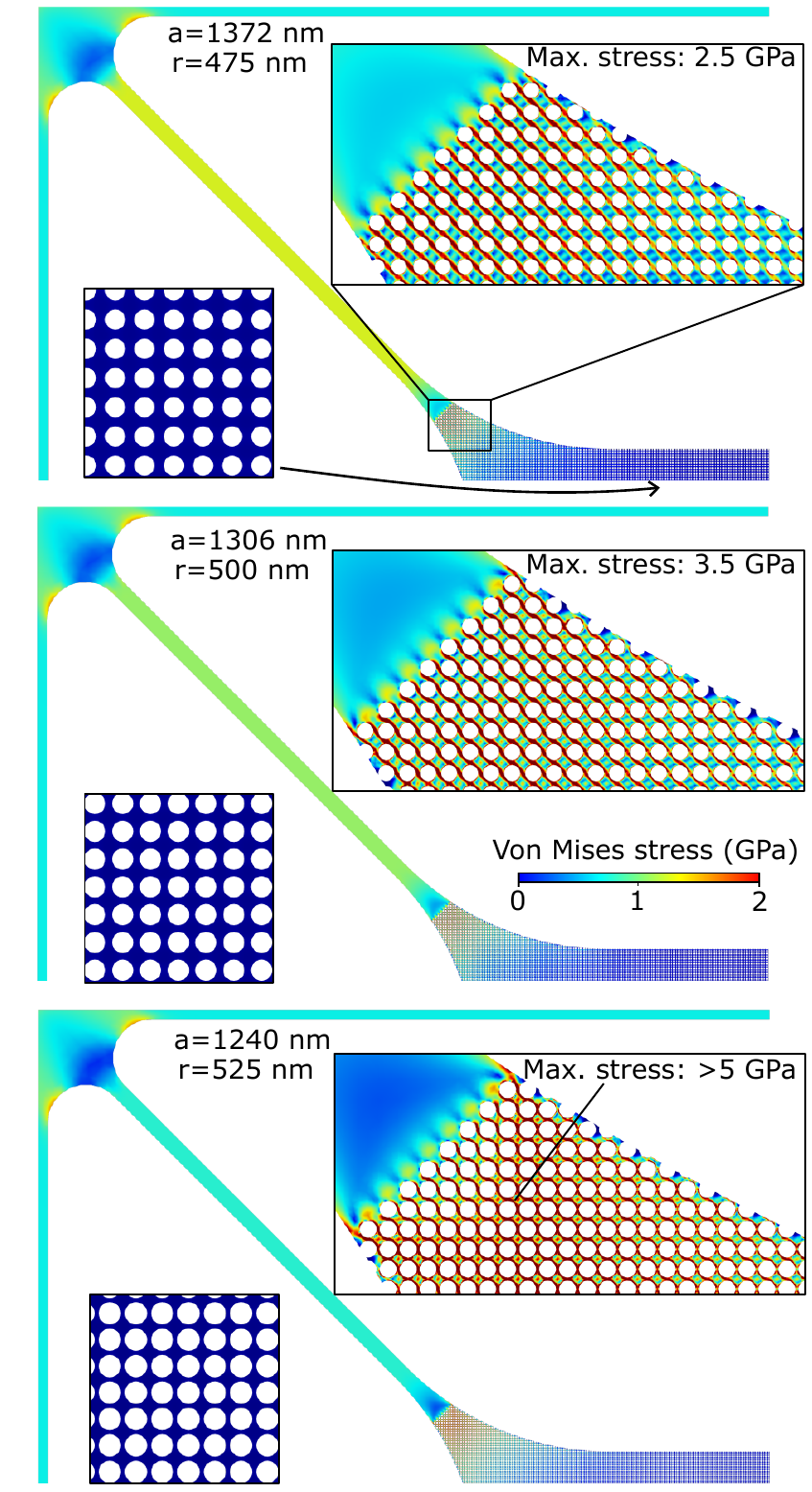}
\caption{Von Mises stress in Si$_3$N$_4$ membrane due to \SI{1}{\giga\pascal} initial stress redistributing upon release of suspended structure, for different photonic crystal parameters. The stress is focused at the start of the photonic crystal, depending on the photonic crystal lattice parameter $a$ and hole size $r$. In the center of the pad, the stress is independent of these parameters. Color scale is the same for all panels.}
\label{figure_stress_redistribution}
\end{figure}

We fabricate the devices from a \SI{1}{\milli\meter} thick Si wafer coated with \SI{100}{\nano\meter} Si$_3$N$_4$ on both sides, which have a \SI{1}{\giga\pascal} tensile pre-stress. When we release the patterned membranes in a dry-release etch step, the stress redistributes. In sweeping the photonic crystal lattice spacing $a$ and hole radius $r$, we found that the yield is reduced below a certain mass ratio and that the membranes broke often at the edge of the photonic crystal. We simulate this stress redistribution in our suspended membrane for three different mass ratios, the lowest and highest one used for the measurements in the main text (Fig.~\ref{figure_stress_redistribution}, top and middle panel) and one where most membranes collapsed (bottom panel). \\

The redistribution of the stress shows a clear pattern: In the center of the membrane pad, the stress is reduced to zero regardless of the photonic crystal parameters. At the edge of the photonic crystal, the stress greatly increased, because the effective width perpendicular to the tensile axis (diagonal in Fig.~\ref{figure_stress_redistribution}, along the length of the tether) is reduced by the photonic crystal holes. For the lattice spacing and hole radius where the yield was low (bottom panel), the stress in this region approaches the yield stress of Si$_3$N$_4$. In the tether itself, the stress is lower the closer we get to this low-yield region. This is likely due to the membrane having less material to pull the tether when the mass ratio of the photonic crystal is low, which results in low tether stress and high stress at the photonic crystal edge. The yield stress of Si$_3$N$_4$ thus limits the achievable mass ratio by sweeping the photonic crystal parameters. This corresponds to the photonic crystal edge being a common point of failure of the devices. \\

In Fig.~\ref{figure_stress_redistribution}, we show the stress distribution depends on the photonic crystal parameters. Since the bending losses of these resonators are governed by dissipation dilution due to stress, we estimate the change in Q from the different stress distributions. We obtain the elongation ($U_\mathrm{el}$) and bending ($U_\mathrm{be}$) energies of the fundamental mode of each resonator by integrating over the domain $S$\cite{Fedorov2019,Yu2012,Shin2022},
\begin{equation}
\begin{aligned}
&U_\mathrm{el} = t \int \left(\sigma_{xx} u_{z,x}^2 + \sigma_{yy} u_{z,y}^2 + \sigma_{xy} u_{z,x}u_{z,y} \right) \mathrm{d}S, \\
&U_\mathrm{be} = \frac{E t^3}{12(1-\nu^2)} \times\\
&\int \!\left( u_{z,xx}^2 \!+\! u_{z,yy}^2 \!+\! 2\nu u_{z,xx}u_{z,yy} \!+\! 2(1 \!-\! \nu)^2 u_{z,xy} \right) \mathrm{d}S.
\end{aligned}
\end{equation}
Here, $\sigma$ is the stress distribution, $u_z$ the out-of-plane resonator displacement and the comma denotes derivative with respect to that coordinate. We use thickness $t = 80$~nm for the Si$_3$N$_4$, which is reduced from the deposited thickness (\SI{100}{nm}) by the etching, $E = 250$~GPa the Young's modulus and $\nu = 0.23$ the Poisson ratio of Si$_3$N$_4$. This ratio of these two energies gives us the enhancement of Q from the intrinsic (bending) $Q_0$ of un-stressed Si$_3$N$_4$ due to dissipation dilution, via
\begin{equation}
Q = Q_0 \left(1 + \frac{U_\mathrm{el}}{U_\mathrm{be}}\right).
\end{equation}
with\cite{Villanueva2014,Shin2022} $Q_0 \simeq 6900 t/100$~nm to normalize it to a Si$_3$N$_4$ thickness of \SI{100}{\nano\meter}.

\begin{figure}
\includegraphics[width = 0.5\textwidth]{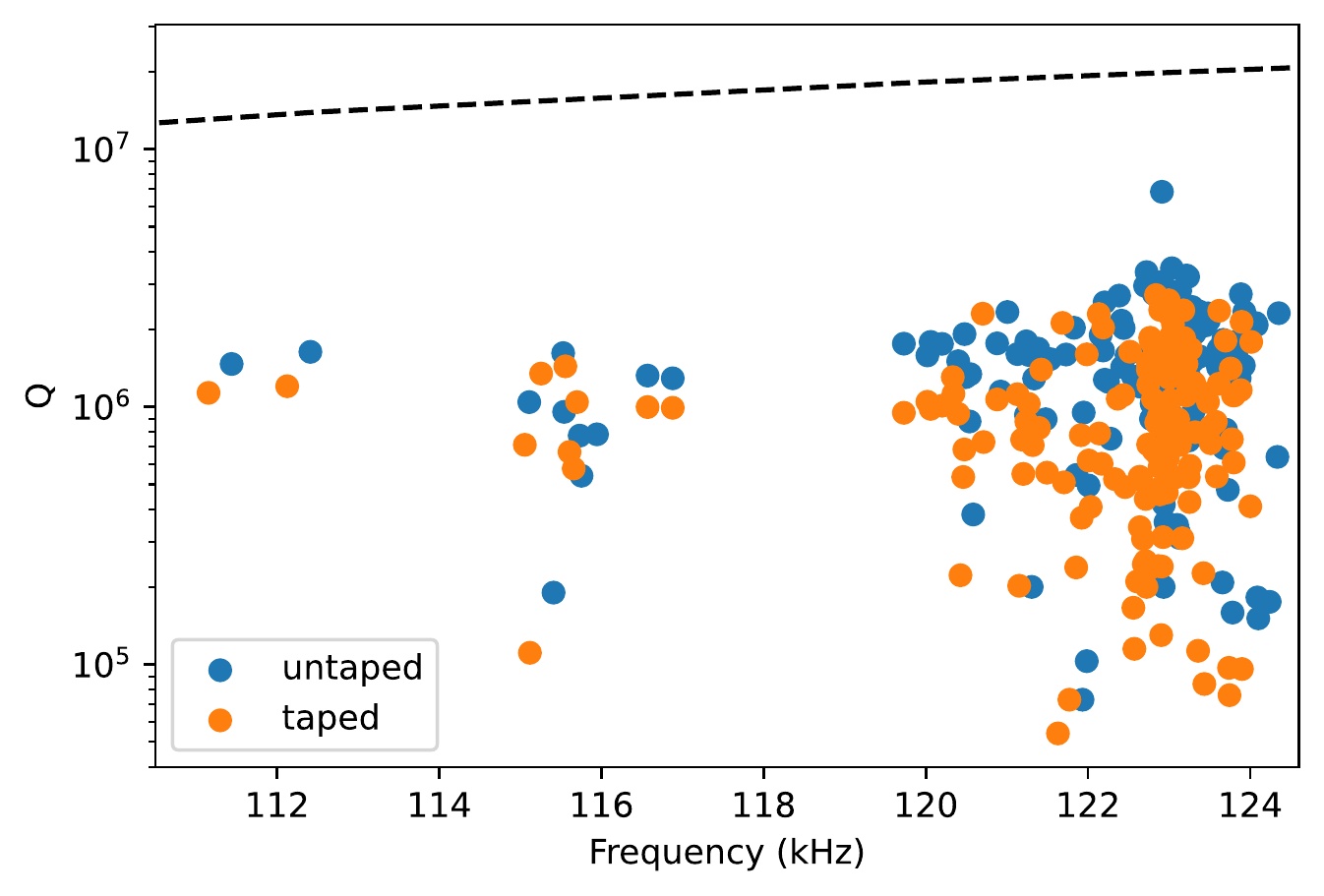}
\caption{Measured Q-factors of the resonators, for the taped and untaped substrate. Black dashed line denotes the bending-loss-limited Q-factor calculated from the model in the text.}
\label{figure_allqs}
\end{figure}

From the simulations of the stress redistribution, we calculate the elongation and bending energies of the fundamental membrane mode, and report the expected enhancement in Q due to the dissipation dilution in Table \ref{Table_Qenhancement}. Over the range of parameters with a high fabrication yield, the enhancement of the Q is not too dissimilar ($20\%$ change between the top two rows). This validates our assumption that changing the photonic crystal parameters is not the dominant factor in any trends in the Q that we see. Furthermore, when we plot all the Q-factors of the resonators in the main text, Fig.~\ref{figure_allqs}, we see that the bending-loss limited Q-factor that follows from the model (dashed black line) is higher than the measured ones, meaning we are likely not in the bending-loss-limited regime. 

\begin{table}
\begin{tabular}{l|lll}
\hline
Parameters & $U_\mathrm{el}$(J) & $U_\mathrm{be}$(J) & $Q/Q_0$ \\
\hline
$a = 1372$~nm & $7.6\cdot10^{-26}$& $2.4\cdot10^{-29}$& 3198\\ 
$r = 475$~nm & & & \\
\hline
$a = 1306$~nm & $4.4\cdot10^{-26}$& $1.7\cdot10^{-29}$& 2628\\ 
$r = 500$~nm & & & \\
\hline
$a = 1240$~nm & $2.3\cdot10^{-26}$& $1.5\cdot10^{-29}$& 1536\\ 
$r = 525$~nm & & & \\
\hline
\end{tabular}
\caption{Q enhancement due to dissipation dilution for the different photonic crystal parameters}
\label{Table_Qenhancement}
\end{table}


\section{Q-reduction by laser heating}
We perform ringdown measurements by driving resonators with a sinusoidal excitation at their resonance frequency, which results in a high velocity detected by the Polytec MSA400 vibrometer. In some of the resonators, a significant velocity can be detected without any driving signal applied to the piezoelectric mounted on the sample holder. Some measurements even show negative Q-factors (i.e. increase in resonator displacement while the piezo shaker is not driven), indicating that there is another drive mechanism present in our system. This drive mechanism is the subject of future work, so we limit ourselves to a brief description and focus on the effect it has on resonator Q-factor.

In short, the driving is associated with strong absorption of the \SI{633}{\nano\meter} laser light due to the photonic crystal structures on our membrane. This leads to heating and can cause optothermal self-oscillation if the laser power is sufficiently high~\cite{Aubin2004}, which can drive the mechanical motion from an unmodulated continuous-wave laser. In our system, this likely happens due to the heating causing thermal expansion, which modulates the stress in the membrane. This effectively creates an optothermal parametric drive mechanism. It is not easy to distinguish the presence of this second driving mechanism, especially if it is weaker than the piezoelectric driving.

There are two properties we can associate with the presence of the heating and optothermal driving mechanism. The first is that the Q-factors measured when there is optothermal driving tend to show a large variance (between directly sequential measurements on the same device). This originates from the fact that the photonic crystal structures are somewhat position-dependent in our membrane due to the stress redistribution after release (previous section). Because of this, the absorption, heating and optothermal driving are position-dependent, and Q-factors measured at slightly different positions might be very different. Empirically, we estimated a variance $>20\%$ between the Q-factors of the three ringdowns performed for each device was a good indicator for the presence of the heating and optothermal driving. Thus we use this as a cutoff, and reject all measurements which show this large variance.

The second property is the presence of integer multiples of our fundamental mode, 'overtones' as named by others~\cite{Yang2019,Yang2021}. These indicate the resonator is in the non-linear regime, even though a ringdown measurement might appear linear. Thus we also reject the measurements that display these overtones. 

\begin{figure*}
\includegraphics[width = \textwidth]{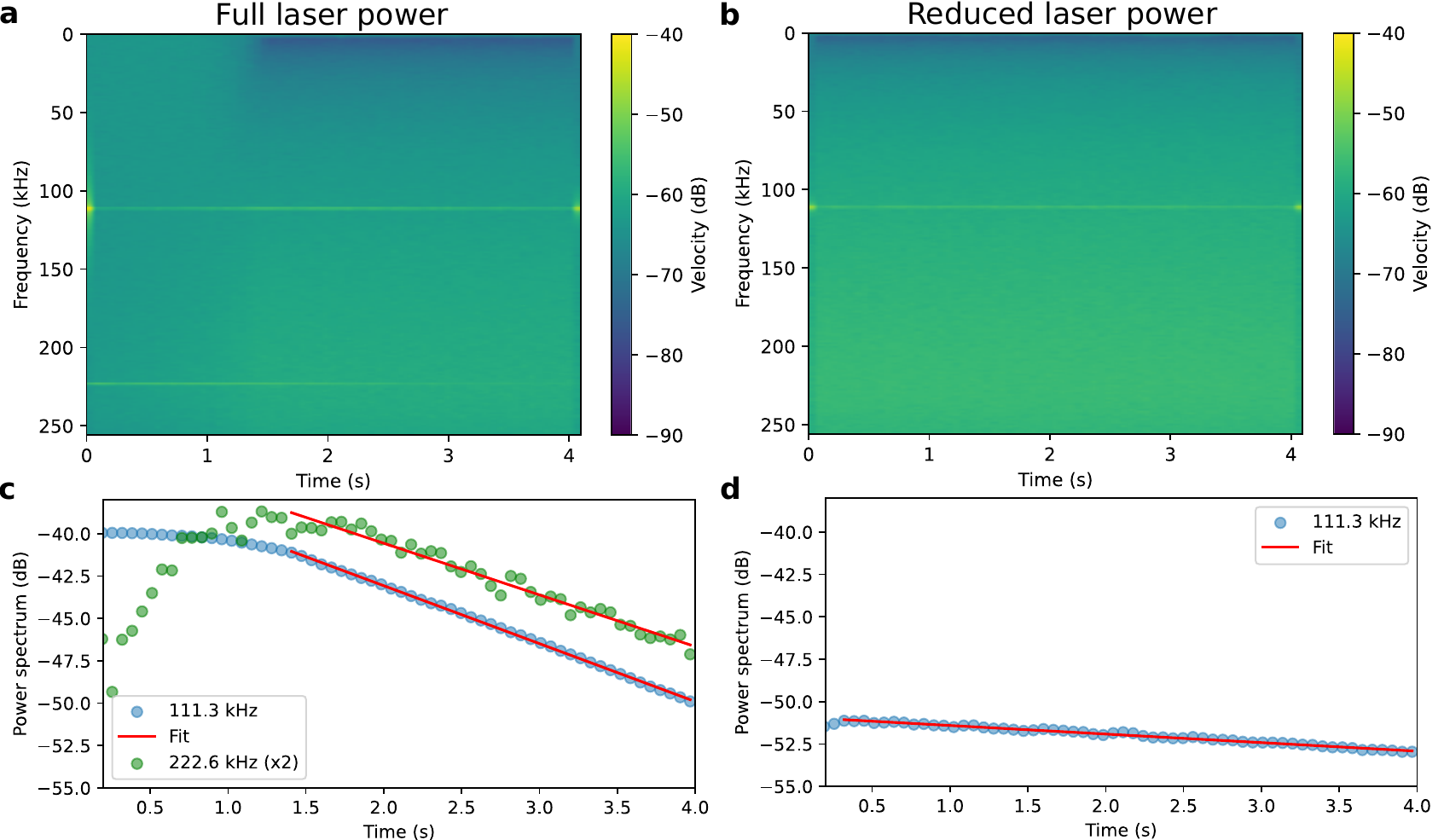}
\caption{\textbf{a,b:} Velocity amplitude spectrum plotted as a function of time for full and reduced laser power. In \textbf{a}, a second mode at exactly double the fundamental mode frequency is visible, which is absent in \textbf{b}. This mode is attributed to the laser driving the resonator into the non-linear regime. \textbf{c,d:} Ringdowns (horizontal time-cuts) of the modes in \textbf{a,b} respectively. The mode at twice the fundamental frequency is weaker, so it has been scaled by a factor of two.}
\label{fig_laser_heating}
\end{figure*}

We show an example of such laser-driven ringdown measurement in Fig.~\ref{fig_laser_heating}, where we measure the same membrane in directly subsequent measurements, but reduce the laser power by adding a neutral density filter (Thorlabs NE506A, 25\% transmission, from \SI{3.6}{\milli\watt} to $\simeq$\SI{1.0}{\milli\watt}) in the beam path for the second measurement. We record the time trace, and perform a Fourier transform with a shifting time window to visualize the change in the velocity spectrum over time. In both Figs.~\ref{fig_laser_heating}\textbf{a} and \textbf{b}, we clearly resolve the fundamental mode, but in Fig.~\ref{fig_laser_heating}\textbf{a}, a second mode at twice the fundamental frequency is present. The presence of this mode demonstrates that the membrane is in the non-linear regime~\cite{Yang2019,Yang2021}. We want to stress that this second mode is not related to the transduction non-linearity common to large displacements in interferometric setups (see e.g. Dolleman et al.\cite{Dolleman2017}), as laser Doppler vibrometers are not affected by these.

We take a horizontal cut of the spectra of Fig.~\ref{fig_laser_heating} to obtain the ringdowns of the modes at the fundamental frequency, and at precisely double the fundamental frequency, which we plot in Fig.~\ref{fig_laser_heating}\textbf{c,d}. By performing a linear fit, we extract the Q-factor of each of these modes, and the Q-factor from the reduced power measurement (\textbf{d}, $Q = 3.83\cdot 10^6$ at \SI{111.3}{\kilo\hertz}) is larger than the Q-factors from the full-power measurement (\textbf{c}, $Q = 0.57\cdot 10^6$ at \SI{111.3}{\kilo\hertz} $Q = 0.63\cdot 10^6$ at \SI{222.6}{\kilo\hertz}). The difference in the operating laser power (cf. heating and optothermal driving)is thus related to a large difference in the Q-factor measured by a ringdown.  

The ringdowns and Q-factors reported supporting the figures in the main text were all taken at the minimum operating laser power of~\SI{1}{\milli\watt}. However, the heating and optothermal driving do not appear equally strong in all membranes, and some resonators are still driven into the non-linear regime even at reduced power. To further reduce the absorption and increase the thermal contact to the substrate, we read out the velocities from the tether foot (closest to the substrate) instead of at the membrane pad. This comes at a cost of reducing the measured velocity, as the tether foot has a much lower motional amplitude than the center of the membrane, reducing the signal-to-noise ratio. This makes it difficult to gauge if we are in the linear regime, as the mode at twice the fundamental frequency might be hidden by the detector noise.

In summary, there is a heating and optothermal driving effect present in some of our membranes, due to the absorption facilitated by the photonic crystal structures. This driving mechanism happens even at the minimum operating power of the setup, and affects the Q-factors obtained from a ringdown. To exclude these effects, we remove the devices from our dataset if they display a large variance ($>20\%$) in the Q-factors measured in directly sequential ringdowns, or if they display the second mode (overtone) in their spectra. 

\section{Resonators coupled via the substrate}
\begin{figure}
\includegraphics[width = 0.5\textwidth]{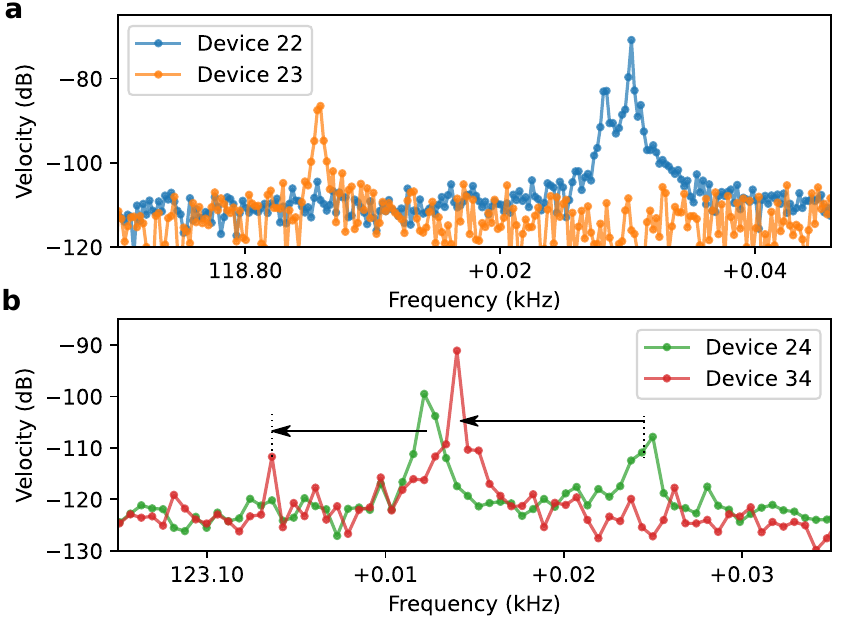}
\caption{\textbf{a} Spectra of the two mechanical resonators coupled resonators in the main text. The split peak of device 22 shows coupling. Device 23 was measured 1 hour later, the frequency decreased over time due to creep. \textbf{b} Spectra of two devices measured in swift succession (20 minutes delay), demonstrating that both devices show the split peak. In this time frame, there is a \SI{9}{\hertz} downwards frequency shift. Arrows identify the shifted peaks, the peak of the mode belonging to the read-out membrane itself is stronger than the one of the coupled membrane.}
\label{figcoupledspectra}
\end{figure}

In the main text, we describe two specific resonators coupled to each other via the substrate. To identify which specific membranes are coupled, we compare the resonance frequencies of the set of five resonators with nominally the same design and frequency. By comparing their resonance frequencies, the resonator reported in the main text (device 22, blue in Fig.~\ref{figcoupledspectra}\textbf{a}) was most likely coupled to device 23 (orange in Fig.~\ref{figcoupledspectra}\textbf{a}). The displayed spectrum of device 22 was taken before the measurements of the main paper, shown in Fig.~5a, while the spectrum of device 23 was taken after those measurements, and thus approximately one hour after the spectrum of device 22 was measured. It follows that there was likely some creep in the Si$_3$N$_4$ that caused a downshift of the spectrum of device 23, meaning it was closer in frequency to device 22 when the coupling was observed. 

As the system is fully passive (i.e. there is not mechanism to actively tune the frequency with) and creep is not reversible, it is difficult to a posteriori verify which two resonators were coupled. However, the reported mechanical modes are the fundamental modes of the trampoline resonators, which do not show splitting in case of imperfections. This excludes the possibility of the split peak originating from the same resonator. Furthermore, we have observed similar coupling in more devices, specifically devices 24 and 34 of a different chip. Here, the spectra were measured in quick succession ($<20$ minutes delay), plotted in Fig.~\ref{figcoupledspectra}\textbf{b}. In both spectra, two peaks are visible, shifted by approximately the same amount. We identify the stronger peak as the one belonging to the read-out membrane, as the peak from the coupled membrane is likely weaker.

Based on the observation of a pair of resonator spectra where both show the double peak, and the quantitative match between the two-coupled-resonator model and the data in the main text, we consider it clear that there is coupling between two discrete modes. By the arguments above, it is most likely that it is coupling between devices 22 and 23. On the chip, these resonators are separated by \SI{1.5}{\milli\meter} (center-to-center distance). \\

In the main text, we use two models to describe coupling between the high-Q resonator and the substrate (Fig.~\ref{figure_model_comparison}\textbf{a}, stacked model), and between two high-Q resonators (coupled model). For the resonator-substrate coupling, the stacked model provides a straightforward match to the physical system: The Si$_3$N$_4$ resonator is on top of the Si substrate on top of the sample holder ('ground'), so any motion of the substrate automatically affects the position of the Si$_3$N$_4$ resonator. Conversely, for the resonator-resonator coupling the positions of the resonators are effectively independent, except for some weak coupling spring that moves energy from one resonator to the other, so the coupled model provides the most straightforward description. 

While the two models appear different, their parameters can be related relatively straightforwardly if the damping is small. We use the parameters from the main text, $m_1, m_2$ for the effective resonator masses and $k_1, k_2$ for their spring constants. The coupled model has additional spring $k_3$ that couples the two resonators, by which we isolate the coupling rate $J$ between the resonators such that $J^2 = \frac{k_3^2}{m_1m_2}$. In the regime of low damping (i.e. no viscous term accompanying $k_3$), we can directly relate the parameters of the stacked model to those of the coupled model via 
\begin{equation}
\begin{aligned}
k_{s,2} &= k_{c,2} + k_{c,3} \\
k_{s,1} &= \frac{k_{c,1} + k_{c,3}}{k_{c,3}^2}k_{s,2}^2 - k_{s,2} \\
m_{s,1} &= \frac{k_{s,2}^2}{k_{c,3}^2} m_{c,1} \\
m_{s,2} &= m_{c,2},
\end{aligned}
\end{equation}
where the subscripts $s,c$ denote the stacked and coupled model parameters respectively. This allows for translation of the coupling strength between the resonators to the coupling strength between resonator and substrate. \\

\begin{figure}
\includegraphics[width = 0.5\textwidth]{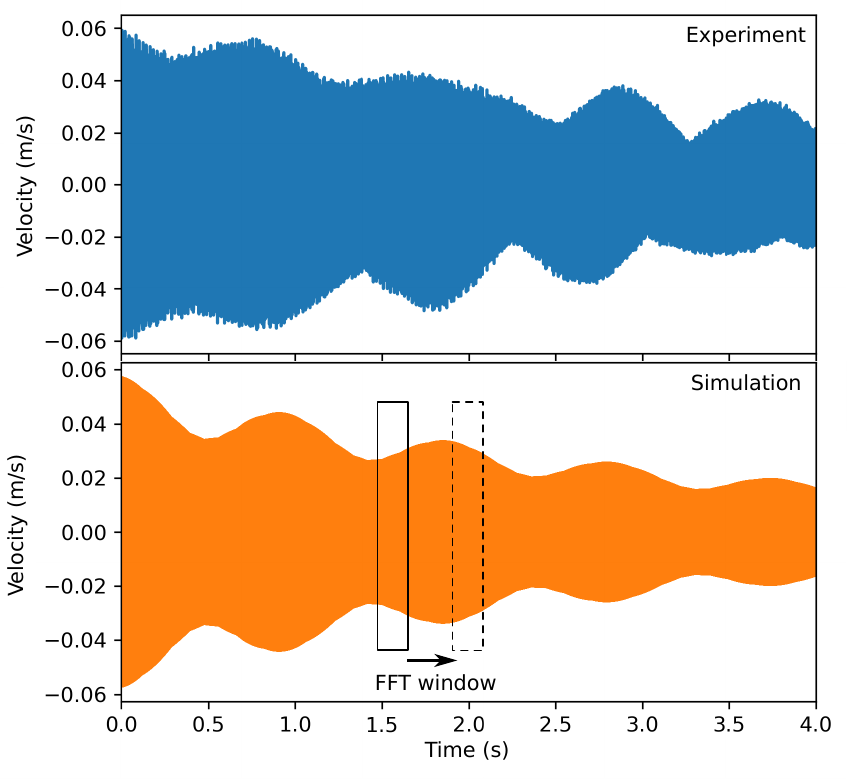}
\caption{Experimentally obtained (top) and simulated (bottom) velocities of resonator ringdown. Performing a Fourier transform with a shifting time window allows extracting the time-dependent amplitude decay that describes the ringdown of a resonator mode(s).}
\label{figure_time_ringdown}
\end{figure}

\begin{figure}
\includegraphics[width = 0.5\textwidth]{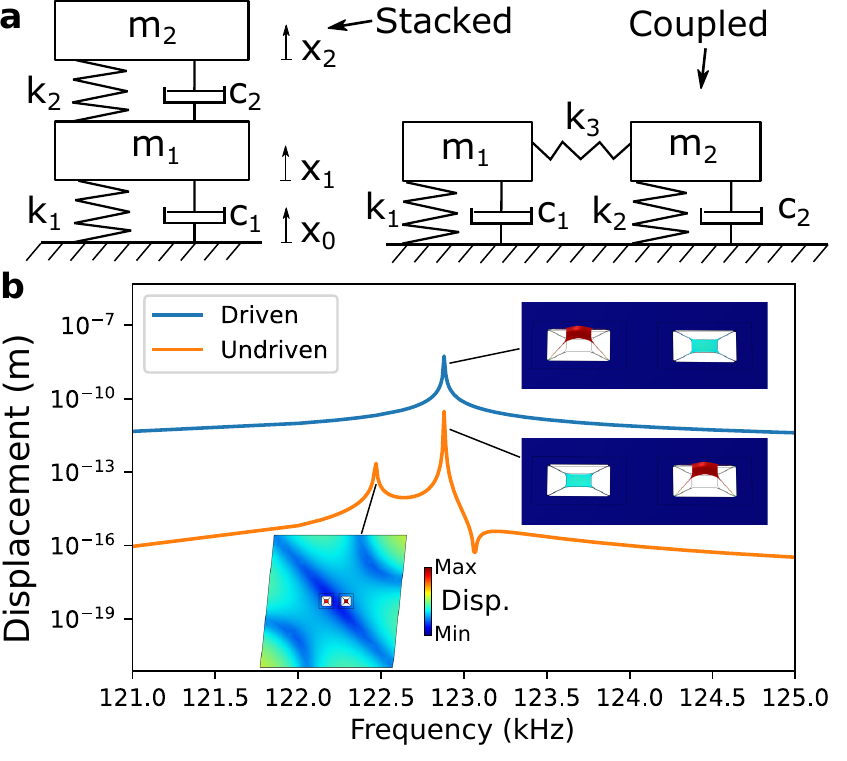}
\caption{\textbf{a:} Schematic of stacked 2-DOF model and of coupled two-resonator model. \textbf{b:} Frequency response of membrane driven with \SI{1}{\pico\newton} harmonic perturbation (blue) and of the membrane coupled through the substrate. Insets show substrate resonance and degenerate resonator modes.}
\label{figure_model_comparison}
\end{figure}

To simulate the coupling between the resonators in the main text, we start from the equations of motion for the resonator positions from the coupled-resonator model,
\begin{equation}
\begin{aligned}
\ddot{x}_1 + \gamma_1 \dot{x}_1 + \omega_1^2 x_1 + J^2 x_2 &= 0 \\
\ddot{x}_2 + \gamma_2 \dot{x}_2 + \omega_2^2 x_2 + J^2 x_1 &= 0.
\end{aligned}
\end{equation}
We can write these as a set of four coupled first-order differential equations (for $[x_1, v_1, x_2, v_2]^T$, the positions and velocities of the two resonators respectively) and numerically integrate them. The resulting velocity of one of the resonators can be extracted and compared to the measured velocity, which we show in Fig.~\ref{figure_time_ringdown}. The envelopes of the two curves match very well.

The resulting time-trace is Fourier-transformed with a shifted time window in exactly the same manner as the measured data, which sacrifices some frequency resolution but allows us to extract the time-dependent behavior of the resonance peaks. This way, we extract a ringdown measurement of a specific mode from the time trace, and the result is shown in the main text.\\

We can further corroborate the existence of coupling between individual resonator on chip by utilizing a finite element model of a substrate with two trampoline resonators, Fig.~\ref{figure_model_comparison}\textbf{b}. The two resonators are meshed identically to ensure their eigenmodes are the same, and spaced \SI{1.5}{\milli\meter} apart on the substrate. We evaluate their coupling by adding a \SI{1}{\pico\newton} out-of-plane harmonic perturbation force in the center of one of the resonators, and tracking the resulting displacement in the center of both resonators. 

For the driven resonator (Fig.~\ref{figure_model_comparison}\textbf{b}, blue curve), the resulting displacement spectrum is sharply peaked around the fundamental mode. For the undriven resonator (orange curve) we see two peaks, one associated with the substrate mode and another associated with the resonator mode, shown in the insets. There is also an antiresonance visible, where the response of the undriven resonator is perfectly out-of-phase with the response of the driven resonator. The peak at resonance means that there is energy transfer from one resonator to the other. At the resonator frequency, the simulated amplitude of the undriven resonator is a factor $\sim$180 smaller than that of the driven resonator, which is not too far off of the ratio $\omega_1/J \approx 140$. This suggests that the coupling rate extracted from the fits is a reasonable match with the FEM simulations.

\end{document}